\input harvmac
%
%
%
%
%
%
%
%
%
%
%
%
%
%

\edef\recoveratcodezqrz{\catcode`\noexpand\@=\the\catcode`\@}

\catcode`\@=11

%

\def\m@ltiplybyscale#1,#2,#3.{%
  \multiply #1 by #2\relax
  \divide #1 by #3\relax
}

\newif\ifAMSTEX@loaded
\newif\ifAMSPPT@loaded

\def\loc@lm@g#1#2#3{%
  %
  \expandafter\ifx\csname amsppt.sty\endcsname\relax
       \AMSPPT@loadedfalse\else\AMSPPT@loadedtrue\fi
  \ifx\amstex@loaded\relax\AMSTEX@loadedtrue\else\AMSTEX@loadedfalse\fi
  %
  %
  \edef\c@ntfive{\the\count5}%
  \count255=#1  
  %
  %
  %
  \ifx\MathTimefontnames\relax
    \message{<Assuming MathTime style font definitions>}%
    %
    \edef\c@untfive{\the\count5}%
    \edef\c@untsix{\the\count6}%
    \edef\c@untseven{\the\count7}%
    \edef\c@unteight{\the\count8}%
    %
    \count5=\count255
    \m@ltiplybyscale\count5,1,2.
    %
    \count6=\count255
    \m@ltiplybyscale\count6,3,5.
    %
    \count7=\count255
    \m@ltiplybyscale\count7,7,10.%
    %
    \count8=\count255
    \m@ltiplybyscale\count8,4,5.
    %
%
    \font\tenrm=\RomanFont scaled \count255
    \font\sevenrm=\RomanFont scaled \count5
    \font\fiverm=\RomanFont scaled \count7
    \textfont0=\tenrm
    \scriptfont0=\sevenrm
    \scriptscriptfont0=\fiverm
%
    \font\teni=MTMI scaled \count255
    \font\seveni=MTMI scaled \count7
    \font\fivei=MTMI scaled \count5
    \ifAMSPPT@loaded
      \font\eighti=MTMI scaled \count8\relax  \skewchar\eighti=45
      \font\sixi=MTMI scaled \count6\relax    \skewchar\sixi=45
    \fi
    \textfont1=\teni
    \scriptfont1=\seveni
    \scriptscriptfont1=\fivei
%
    \font\tensy=MTSY scaled \count255
    \font\sevensy=MTSY scaled \count7
    \font\fivesy=MTSY scaled \count5
    \textfont2=\tensy
    \scriptfont2=\sevensy
    \scriptscriptfont2=\fivesy
%
    \font\tenex=MTEX scaled \count255
    \font\sevenex=MTEX scaled \count7
    \textfont3=\tenex
    \scriptfont3=\sevenex
%
    \font\tenit=\ItalicFont scaled \count255
    \font\sevenit=\ItalicFont scaled \count7
    \textfont\itfam=\tenit
    \scriptfont\itfam=\sevenit
%
    \font\tensl=\SlantFont scaled \count255
    \textfont\slfam=\tensl
%
    \font\tenbf=\BoldfaceFont scaled \count255
    \textfont\bffam=\tenbf
%
    \font\tentt=\TypewriterFont scaled \count255
    \textfont\ttfam=\tentt
    %
    \count5=\c@untfive
    \count6=\c@untsix
    \count7=\c@untseven
    \count8=\c@unteight
  %
  \else    
    \message{<Assuming Computer Modern style font definitions>}%
    \font\tenrm=cmr10 scaled \count255
    \font\sevenrm=cmr7 scaled \count255
    \font\fiverm=cmr5 scaled \count255
    \ifAMSPPT@loaded 
      \font\eightrm=cmr8 scaled \count255
      \font\sixrm=cmr6 scaled \count255
    \fi

    \textfont0=\tenrm
    \scriptfont0=\sevenrm
    \scriptscriptfont0=\fiverm
    \font\teni=cmmi10 scaled \count255
    \font\seveni=cmmi7 scaled \count255
    \font\fivei=cmmi5 scaled \count255
    \ifAMSPPT@loaded
      \font\eighti=cmmi8 scaled \count255
      \font\sixi=cmmi6 scaled \count255
    \fi

    \textfont1=\teni
    \scriptfont1=\seveni
    \scriptscriptfont1=\fivei
    \font\tensy=cmsy10 scaled \count255
    \font\sevensy=cmsy7 scaled \count255
    \font\fivesy=cmsy5 scaled \count255
    \ifAMSPPT@loaded
      \font\eightsy=cmsy8 scaled \count255
      \font\sixsy=cmsy6 scaled \count255
    \fi

    \textfont2=\tensy
    \scriptfont2=\sevensy
    \scriptscriptfont2=\fivesy
    \font\tenex=cmex10 scaled \count255
    \textfont3=\tenex
    \scriptfont3=\tenex
    \scriptscriptfont3=\tenex
  \ifAMSPPT@loaded
      \font\sevenex=cmex7 scaled \count255
      \font\eightex=cmex8 scaled \count255
      \textfont3=\tenex
      \scriptfont3=\sevenex
      \scriptscriptfont3=\sevenex
    \fi

    \font\tenit=cmti10 scaled \count255
    \ifAMSPPT@loaded
      \font\sevenit=cmti7 scaled \count255
      \scriptfont4=\sevenit
      \scriptscriptfont4=\sevenit
      \font\eightit=cmti8 scaled \count255
    \else
      \ifAMSTEX@loaded
        \scriptfont4=\tenit
        \scriptscriptfont4=\tenit
      \fi
    \fi

    \textfont\itfam=\tenit
    \font\tensl=cmsl10 scaled \count255
    \ifAMSTEX@loaded
      \scriptfont5=\tensl
      \scriptscriptfont5=\tensl
    \fi
    \ifAMSPPT@loaded
      \font\eightsl=cmsl8 scaled \count255
    \fi

    \textfont\slfam=\tensl
    \font\tenbf=cmbx10 scaled \count255
    \ifAMSPPT@loaded
      \font\eightbf=cmbx8 scaled \count255
      \font\sixbf=cmbx6 scaled \count255
    \fi

    \textfont\bffam=\tenbf
    \scriptfont\bffam=\sevenbf
    \scriptscriptfont\bffam=\fivebf
    \font\tentt=cmtt10 scaled \count255
    \ifAMSPPT@loaded
      \font\eighttt=cmtt8 scaled \count255
    \fi

    \textfont\ttfam=\tentt
  \fi  
%
%
  \m@ltiplybyscale\baselineskip,#2,#3.\relax
  \m@ltiplybyscale\normalbaselineskip,#2,#3.\relax
  \m@ltiplybyscale\normallineskip,#2,#3.\relax
%
%
  \m@ltiplybyscale\thinmuskip,#2,#3.\relax
  \m@ltiplybyscale\medmuskip,#2,#3.\relax
  \m@ltiplybyscale\thickmuskip,#2,#3.\relax
%
%
  \m@ltiplybyscale\smallskipamount,#2,#3.\relax
  \m@ltiplybyscale\medskipamount,#2,#3.\relax
  \m@ltiplybyscale\bigskipamount,#2,#3.\relax
  \m@ltiplybyscale\jot,#2,#3.\relax
%
%
  \m@ltiplybyscale\parindent,#2,#3.\relax
  \m@ltiplybyscale\parskip,#2,#3.\relax
%
%
%
  \normalbaselines
  \rm
}

\let\localmag\loc@lm@g

\recoveratcodezqrz


\newcount\figno
\figno=0
\def\fig#1#2#3{
\par\begingroup\parindent=0pt\leftskip=1cm\rightskip=1cm\parindent=0pt
\baselineskip=11pt
\global\advance\figno by 1
\midinsert
\epsfxsize=#3
\centerline{\epsfbox{#2}}
\vskip 12pt
{\bf Fig. \the\figno:} #1\par
\endinsert\endgroup\par
}
\def\figlabel#1{\xdef#1{\the\figno}}
\def\encadremath#1{\vbox{\hrule\hbox{\vrule\kern8pt\vbox{\kern8pt
\hbox{$\displaystyle #1$}\kern8pt}
\kern8pt\vrule}\hrule}}

\overfullrule=0pt

\Title{MIT-CTP-2739}
{\vbox{\centerline{Absorption of partial waves by three-branes 
 }}}
\smallskip
\smallskip
\centerline{Samir D. Mathur\foot{E-mail: me@ctpdown.mit.edu}}
\smallskip
\centerline{and}
\smallskip
\centerline{Alec Matusis\foot{E-mail: alec\_m@ctp.mit.edu}}
\bigskip

\centerline{\it Center for Theoretical Physics}
\centerline{\it Massachusetts Institute of Technology}
\centerline{\it Cambridge, MA 02139, USA}
\bigskip

\medskip

\noindent

We study the absorption of a class of fields in the geometry produced by
 extremal three-branes. We consider fields that do not mix with the ten-dimensional
 graviton. For these fields we solve the wave equations and find the  
absorption probabilities for all partial waves at leading order in the
 energy. We note that in some of these cases one needs an `intermediate'
 region which  interpolates between flat Minkowski space at infinity and
 the AdS geometry near the branes.

\Date{May, 1998}

\overfullrule = 0pt
\newsec{Introduction}

There has been renewed interest in the study of the background produced 
by 3-branes, due to their conjectured connection to large N Yang-Mills 
theories in four spacetime dimensions \ref\pol0{A. Polyakov, `String 
Theory and Quark Confinement', hep--th/9711002}\ref\mald{J. Maldacena,
 `The Large $N$ Limit of Superconformal
Theories and Supergravity', hep--th/9711200}\ref\polyakov{S.S. Gubser,
 I.R. Klebanov and A.M. Polyakov,
`Gauge Theory Correlators from Non--critical String Theory',
hep--th/9802109}\ref\witten{E. Witten, `Anti--De Sitter
Space and Holography', hep--th/9802150}\ref\hash{Steven S. Gubser, 
Akikazu Hashimoto, Igor R. Klebanov and 
Michael Krasnitz,
`Scalar Absorption and the Breaking of the World Volume
Conformal Invariance', hep--th/9803023}.
 Close to the branes, the geometry   reduces to an an anti de-Sitter 
($AdS_5$) metric times an $S^5$, with a self-dual five form field strength
 producing the required curvature. It is conjectured that 10 dimensional 
IIB supergravity on this compactification is dual to large N Yang-Mills 
on the four dimensional boundary of the 5-dimensional AdS space.

Away from the immediate vicinity of the branes, the metric is not of the
 AdS form, and in fact far from the branes the spacetime approaches the
 Minkowski vacuum. The following process is of interest  in this complete
geometry. One imagines a low energy quantum incident on the branes, 
starting at spatial infinity ($r\rightarrow\infty$). Some part of this
 wave tunnels through to the `horizon region' close to the branes 
$r\rightarrow 0$, where it then propagates freely inwards towards 
the branes, and this part of the wave may be considered as absorbed
 by the branes. The remainder of the wave is scattered back to 
$r\rightarrow\infty$. One can compute the absorption probability 
for each spherical partial wave  around the branes, or convert 
this to a cross section for absorption of different fields by the branes. 

The above calculation is closely related to the connection between
supergravity and Yang-Mills. In this connection one needs to find the 
coupling between supergravity modes and operators in the Yang-Mills, 
such that the former act as sources for the operators in the latter.
 Where would this set of couplings come from? Some constraint may arise
 from superconformal symmetry \ref\ferrara{Sergio Ferrara, Christian 
Fronsdal and Alberto Zaffaroni, `On N=8 
Supergravity on $AdS_5$ and N=4 Superconformal Yang-Mills theory', 
hep-th/9802203}, but this is not likely to be enough by itself. In 
earlier work it had been conjectured that the absorption
of supergravity modes by the three branes could be obtained from 
coupling the branes to their background through a Born-Infeld type
 action. Such a coupling gives exact agreement for the three-charge 
black hole \ref\grey{Sumit R. Das and Samir D. Mathur,
'Comparing decay rates for black holes and D-branes', 
Nucl.Phys. B478 (1996) 561}, and for some modes of the dilaton in the
 case of 3-branes \ref\kleb{Igor R. Klebanov, `World Volume Approach 
to Absorption by Non-dilatonic Branes',  Nucl.Phys. B496 (1997) 231}\ref\gubs{S.S. 
Gubser, I.R. Klebanov and A.A. Tseytlin, `String Theory and Classical 
Absorption by Threebranes',  Nucl.Phys. B499 (1997) 217}\ref\gubs1{Steven S. 
Gubser and Igor R. Klebanov,
`Absorption by Branes and Schwinger Terms in the World
Volume Theory', hep--th/9708005}. 

In carrying out the classical calculation for absorption one notes that 
there are three regions of spacetime that one encounters in general. The
 region of small $r$ is the AdS region, but this may not connect directly
 to the free propagation on flat space at $r\rightarrow\infty$. The 
calculation of the absorption probability in some cases requires matching
 the solutions for small $r$ and for large $r$ by using an `intermediate
 region'. We say that this region is `needed' in those cases where  the
 solutions in the `outer' and `inner' (AdS) regions do not  match on to 
each other as functions of $r$.

In this paper we carry out this calculation for a class of massless fields
 of 10-dimensional supergravity propagating in the background produced by
 3-branes. We consider fields that do not mix with the perturbations of
 the 10-dimensional gravitational field.
For the fields that we study we consider all partial wave components, and
 construct solutions to the wave equation at  leading order in energy. In
 all cases but one the fields separate out from each other, and for these
 cases we calculate the absorption probability for the incident quanta. 

It is hoped that this analysis will shed light on the connection between 
supergravity and Yang-Mills, and also on how one might make a microscopic
 model to study absorption by the black 3-brane as well as by other black
 branes and black holes. This issue has been studied recently in 
\ref\dealw{S.P. de Alwis, `Supergravity the DBI Action and Black Hole Physics',
 hep--th/9804019}\ref\das{Sumit R. Das and Sandip P. Trivedi, `Three Brane 
Action and The Correspondence Between N=4 Yang Mills Theory and Anti De Sitter
 Space`, hep--th/9804149}.

\newsec{Basic relations }

\subsec{Field equations and separation of variables}

We will follow for the most part the notations in \ref\van{H. J. Kim, L.J. 
Romans, and P. van Nieuwenhuizen,
`The Mass Spectrum Of Chiral $N=2$ $D=10$ Supergravity on $S^5$',
Phys. Rev. D32 (1985) 389.}. In this paper the authors had studied 
linearised perturbations of type IIB supergravity about $AdS_5\times 
S^5$. While our geometry has this form only very close to the 3-branes,
 we have spherical symmetry at all $r$ and the eigenvalues of the Hodge-de 
Rham operator on $S^5$  can be taken from this reference. The computation of
 these eigenvalues is carried out in detail in \ref\eastvan{A. Eastaugh and
 P. van Nieuwenhuizen, `Harmonics and Spectra on General Coset Manifolds', 
ITP-SUNY-Stony Brook preprint.}. 

The field equations are
\eqn\zone{R_{\hat\mu\hat\nu}=
-{1\over 6}F_{\hat\mu\hat\rho\hat\sigma\hat\tau\hat\kappa}
F_{\hat\nu}^{\hat\rho\hat\sigma\hat\tau\hat\kappa} }
\eqn\ztwo{F_{\hat\mu\hat\nu\hat\rho\hat\sigma\hat\tau}=
{1\over 5!}\epsilon_{\hat\mu\hat\nu\hat\rho\hat\sigma
\hat\tau\hat\mu'\hat\nu'\hat\rho'\hat\sigma'\hat\tau'}
F^{\hat\mu'\hat\nu'\hat\rho'\hat\sigma'\hat\tau'} }
\eqn\zthree{D^{\hat\mu}\partial_{[\hat\mu}A_{\hat\nu\hat\rho]}=
-{2i\over 3}\dot F_{\hat\nu\hat\rho\hat\sigma\hat\tau\hat\kappa}D^{\hat\sigma}
A^{\hat\tau\hat\kappa} }
\eqn\zfour{D^{\hat\mu}\partial_{\hat\mu}B=0}
In terms of the fields of 10 dimensional supergravity,
\eqn\zfive{A_{\hat\mu\hat\nu}=B^{NSNS}_{\hat\mu\hat\nu}+
iB^{RR}_{\hat\mu\hat\nu} }
The field $B$ is a complex scalar describing the dilaton and the 
RR scalar. We study linearized perturbations around the background
 given by a collection of coincident 3-branes.  
The perturbations can be separated into two classes: those that 
involve or couple to the fluctuations of the 10 dimensional metric,
 and those that decouple from these metric perturbations. We will 
study only the latter perturbations in this paper. We expand the 
fields around their background values. In the following the background
 value of the fields is denoted by a dot above the symbol giving the field. 
The coordinates
on the $S^5$ surrounding the branes are called $y^\alpha$. The spatial
 coordiantes long the brane are called $x^a, ~ a=1,2,3$. The coordinates
 $t,r,x^a$ are collectively denoted  by $x^\mu$. (Here $r$ is a radial 
coordinate; the branes are located at $r=0$.) These  coordinates $x^\mu$
 describe the $AdS_5$ geometry in the region close to the 3-branes.  
The 10-dimensional coordinates carry a `hat'. 
The $\epsilon$ symbol is a tensor with
\eqn\zs{\epsilon_{12r03\alpha\beta\gamma\delta\epsilon}=\sqrt{- g}}
with $\alpha\beta\gamma\delta\epsilon$ giving a frame with positive orientation
 on $S^5$. 
 
Note: \quad We will call a field a p-form if it has p indices of the kind $\mu$.
 This follows the language that is appropriate to studying the $AdS_5\times S^5$
 geometry in the region close to the branes.\foot{In earlier work with 3-branes 
sometimes a different language was used. There one  regarded the three space 
directions along the brane as `internal' coordinates, and then the remaining 
7-dimensional spacetime has coordinates $t,r$ and the five coordinates of type
 $\alpha$ on the $S^5$. In that language a p-form would have p indices from 
these seven possibilities.}

The metric is negative for the coordinate $t$ and positive for the other nine.
 The symbol $[~~]$ on subscripts denotes antisymmetrisation with strength 
one, i.e. $A_{[\alpha\beta]}={1\over 2}(A_{\alpha\beta}-A_{\beta\alpha})$.

The linearised perturbations around the bachground can be expanded as \van\

\eqn\zseven{A_{\mu\nu\alpha\beta}=\dot A_{\mu\nu\alpha\beta} +
a_{\mu\nu\alpha\beta}  }
\eqn\zeight{a_{\mu\nu\alpha\beta}=\sum_{I_{10}}   b_{\mu\nu}^{I_{10}}(x) Y^{I_{10}}_{[\alpha\beta]}(y) }

\eqn\zel{A_{\mu\nu}=\sum_{I_1} a_{\mu\nu}^{I_1}(x)Y^{I_1}(y)  }
\eqn\zell{A_{\mu\alpha}=\sum_{I_5} [a_{\mu}^{I_5}(x)Y^{I_5}_\alpha(y)  
+a_{\mu}^{I_1}D_\alpha Y^{I_1}(y)] }
\eqn\zelp{A_{\alpha\beta}=\sum_{I_{10}} [a_{\mu}^{I_{10}}(x)
Y^{I_{10}}_{[\alpha\beta]}(y)  +a^{I_5}D_{[\alpha }Y^{I_5}_{\beta]}(y) ]}
\eqn\ztw{B=\sum_{I_1} B^{I_1}(x)Y^{I_1}(y) }

Here $Y^{I_1}$ are the spherical harmonics appropriate to scalar functions. 
 Similarly $Y^{I_5}_\alpha$ are the harmonics with one vector index along the
 $S^5$, and $Y^{I_{10}}_{[\alpha\beta]}$ are the harmonics for an antisymmetric
tensor on the $S^5$. 

The part $a^{I_5}$ in \zelp\ and the part $a^{I_1}_\mu$ in \zell\ can be set 
to zero after a gauge transformation. The gauge transformation gives 
\eqn\zc{\delta A_{\alpha\beta}=\partial_\alpha\Lambda_\beta-\partial_\beta
\Lambda_\alpha, ~~\delta A_{\mu\alpha} =\partial_\mu\Lambda_\alpha-
\partial_{\alpha}\Lambda_\mu}
and we set
\eqn\zb{\Lambda_\beta=-{1\over 2}a^{I_5} Y^{I_5}_{\beta}(y),~~
\Lambda_\mu=a_{\mu}^{I_1}Y^{I_1}(y)} 

The background has the metric
\eqn\zr{ds^2=H^{-{1\over 2}}[-dt^2+dx^a dx^a]+H^{{1\over 2}}
[dr^2+r^2d\Omega_5^2] }
where 
\eqn\zt{H=1+{R^4\over r^4} }

The background also has a 5-form field strength of the 4-form 
antisymmetric tensor potential:
\eqn\zy{F_{12r03}=H^{-2}R^4r^{-5} }
We have a corresponding field strength for all five indices of
 $F$ on $S^5$, given through \ztwo .

We will need to compute the the following quantity
\eqn\zq{\eqalign{(n+1)D^\alpha & [D_{[\alpha}B_{\beta_1\dots \beta_n]}]\cr
&=(n+1)g_{\beta_1\beta'_1}\dots g_{\beta_n\beta'_n} {1\over \sqrt{-g}} \partial_\alpha[\sqrt{-g}\partial_{[\alpha' }B_{\beta''_1\dots\beta''_n]}g^{\alpha\alpha'}g^{\beta''_1\beta'_1}\dots
 g^{\beta''_n\beta'_n}]\cr
& \equiv (\triangle_y B)_{\beta_1\dots\beta_n}  \cr }}

\eqn\zten{\triangle_y Y^k_{[\alpha\beta]}= - {1\over \tilde r^2}(k+2)^2
 Y^k_{[\alpha\beta]} , ~~k=1,2,\dots }
\eqn\zten{\triangle_y Y^k_{\alpha}= - {1\over \tilde r^2}(k+1)(k+3) Y^k_{\alpha}
 , ~~k=1,2,\dots  }
\eqn\zten{\triangle_y Y^k= - {1\over \tilde r^2}k(k+4) Y^k , ~~k=0, 1,\dots }
Here $\tilde r$ is the proper radius of the sphere on which the operator 
$\triangle_y$ acts:
\eqn\zw{\tilde r=rH^{{1\over 4}} }

The functions $Y^k_{[\alpha\beta]}$ can be split into two sets under the 
action of
\eqn\ze{*DY^{k, \pm}_{[\alpha\beta]}\equiv \epsilon_{\alpha\beta}{}^{\gamma\delta\epsilon}\partial_\gamma 
Y_{[\delta\epsilon]}=\pm 2i (k+2) Y^{k, \pm}_{[\alpha\beta]} }
where $\epsilon_{\alpha\beta\gamma\delta\epsilon}$ is the volume 
form on the unit 5-sphere.

\subsec{Some mathematical relations}

We consider perturbations of the form
\eqn\za{\phi(t,r,y) = \phi(r)e^{-i\omega t} Y(y) }
We will construct the solutions for $\phi(r)$ 
 by dividing the radial coordinate $r$ into three regions:

(a)\quad The outer region, given by $r>r_1$ for some choice of $r_1$ 
satisfying $
{r_1\over R}>>1$ but also $\omega r_1<<1$. [We are working to leading 
order in $\omega$, so we can always assume that $\omega$ is as small 
as we wish.] In the `outer part' of this outer region, we will have 
$\omega r>>1$, while in the `inner part' of this outer region we will
 have $\omega r<<1$. 

(b)\quad The intermediate region, given by $r_1>r>r_2$, where
 ${r_2\over R}<<1$, but 
\eqn\zu{v(r_2)<<1}
\eqn\zi{ v \equiv {\omega R^2\over r} }
In this region we can ignore $\omega$ in the equation.

(c)\quad The inner region, given by $r<r_2$. The region $r\rightarrow 0$
 corresponds to
$v\rightarrow\infty$. A wave propagating to $v\rightarrow\infty$ is 
considered to be absorbed into
the branes. The boundary condition we impose in the computation of 
cross sections
is that there be no
wave propagating from $v=\infty$ towards smaller $v$.

In the outer region the wave equations turn out to take the form
\eqn\six{\phi_{,rr}+A_1r^{-1}\phi_{,r}+[\omega^2-{B_1\over r^2}]\phi=0}
The solution is
\eqn\tw{\phi=C_1~r^{{1\over 2}[1-A_1]}~ J_{\sqrt{B_1+{1\over 4}(A_1-1)^2}}
~(\omega r) +
 C_2~r^{{1\over 2}[1-A_1]}~N_{\sqrt{B_1+{1\over 4}(A_1-1)^2}}~(\omega
r) }

In the inner region the equations take the form
\eqn\onefourt{\phi_{rr}+A_2r^{-1}\phi_{,r}+{\omega^2R^4\over r^4}
\phi-B_2r^{-2}\phi=0}
which is equivalent to
\eqn\onetwenty{\phi_{,vv}+(2-A_2)v^{-1}\phi_{,v}+[1-{B_2\over v^2}]\phi=0}

The solutions are
\eqn\onetwtwo{\phi=C_5 v^{{1\over 2} (A_2-1)}  J_{\sqrt{B_2+{1\over 4}
(1-A_2)^2}}(v) 
+ C_6 v^{{1\over 2} (A_2-1)} N_{\sqrt{B_2+{1\over 4}(1-A_2)^2}}(v) }

For $v\rightarrow\infty$ we have
\eqn\zf{J_k(v)+iN_k(v)\sim e^{iv}, ~~J_k(v)-iN_k(v)\sim e^{-iv} }
Thus to compute the absorption probability we will take in \onetwtwo\ 
a solution with
\eqn\zg{C_5=1,~~ C_6=i}

In the outer region at $r\rightarrow \infty$, we have
\eqn\zf{J_k(\omega r)+iN_k(\omega r)\sim e^{i\omega r}, ~~J_k(\omega r)
-iN_k(\omega r)\sim
e^{-i\omega r} }

Thus from a solution of the form \tw\ we find for the probability of 
absorption of the spherical wave
\eqn\zg{{\cal P}=1-|{C_1-iC_2\over C_1+iC_2}|^2  }

\subsec{Scattering matrices}

We will write out the solution for each field in each of the three
 regions mentioned above. In may cases it is not essential
to introduce the intermediate region to solve the equation, but we 
introduce this region in all the cases since one of our goals is to 
compare the behavior in this region of the different fields. We also 
wish to see the relation between the power law solutions that arise 
here with the dimensions of operators in the Yang-Mills or Born-Infeld
 descriptions of the 3-branes. 

If one wants to compute the absorption probabilities, one can simplify 
the calculation by using a comparison of fluxes at $r\rightarrow \infty$ 
and $r\rightarrow 0$ \ref\maldastrom{J. Maldacena and A. Strominger, 
`Black hole greybody factors and D-brane spectroscopy', Phys. Rev. D55 
(1997) 861.}. In this case one does not need to compute both the solutions 
to the equation in each region, if one wants the result at leading order
 in the energy. We choose to instead to compute both solutions in each
 region, since from this we can extract the result to the following more
 general question. We can imagine the wave to be incident either from 
$r\rightarrow \infty$ or from $r\rightarrow 0$. The final waveform will have
 components that travel in each of these two directions. 
Thus there is a $2\times 2$ scattering matrix describing the 
scattering/tunneling through the intermediate region. We do not
 write this matrix explicitly, but it can be read off from the solutions in 
each case.

\bigskip
\bigskip
\newsec{Absorption of the dilaton-axion}

The equation \zfour\ with the ansatz \za\ becomes
\eqn\onethreep{r^{-5} \partial_r[r^5B_{,r}]+\omega^2 HB
-k(k+4)r^{-2}B=0}

\subsec{Outer region}

Here the equation is
\eqn\onenine{B_{,rr}+5B_{,r}+[\omega^2-{k(k+4)\over r^2}]B=0}
The solution is
\eqn\oneten{B=C_1r^{-2}J_{(k+2)} (\omega r) + C_2r^{-2}N_{(k+2)} (\omega r) }

\subsec{Intermediate region}

Dropping the $\omega^2$ term, we have
\eqn\onefour{B_{rr}+5r^{-1}B_{,r}-k(k+4)r^{-2}B=0}
The solution is
\eqn\oneonew{B=C_3r^k+C_4r^{-k-4} }

Thus in this case we see that the equation does not have a nontrivial
intermediate region.

\subsec{Inner region}

Here the equation reduces to
\eqn\onefourt{B_{rr}+5r^{-1}B_{,r}+{\omega^2R^4\over r^4}B-k(k+4)r^{-2}B=0}
The solutions are
\eqn\onetwtwo{B=C_5 v^2 J_{(k+2)}(v) + C_6 v^2 N_{(k+2)}(v) }

\bigskip
\bigskip
\newsec{Scalar from the antisymmetric two form}

The field equation arises from \zthree , with the free indices taken to 
lie along the $S^5$:
\eqn\threethreep{r^{-1}\partial_r[H^{-1}ra_{\alpha\beta,r}
]+\omega^2a_{\alpha\beta}-(k+2)^2r^{-2}H^{-1}a_{\alpha\beta}
\mp   4H^{-2}R^4r^{-6}(k+2)a_{\alpha\beta}=0}
Here the sign $\pm$ corresponds to the sign  in the spherical harmonic
 $Y^{k,\pm}_{[\alpha\beta]}$ (eq. \ze ).

\subsec{Outer region}

The equation is
\eqn\threeseven{a_{\alpha\beta,rr}+r^{-1}a_{\alpha\beta,r}+
[\omega^2-{(k+2)^2\over r^2}]
a_{\alpha\beta}=0 }

The solutions are
\eqn\threeeight{a_{\alpha\beta}=C_1J_{(k+2)}(\omega r)+ C_2N_{(k+2)}
(\omega r)}

\subsec{Intermediate region}

The equation is
\eqn\threefour{\phi_{,rr}+{1\over r}{[r^4+5R^4]\over [r^4+R^4]}\phi_{,r}
-{(k+2)\over r^{2}} {[(k+2)r^4+(k+2\pm 4)R^4]\over [r^4+R^4]}   \phi=0}

The solution is
\eqn\threesixp{a_{\alpha\beta}=C_3r^{\pm(k+2)}+C_4r^{\mp(k+2)}[1+{(k+2)
\over (k+2\pm 2)}{R^4\over
r^4}] }

Thus in this case we have a nontrivial intermediate region.

\subsec{Inner region}

The equation here is
\eqn\threethreeppp{a_{\alpha\beta,rr}+5r^{-1}a_{\alpha\beta,r}+
[\omega^2R^4r^{-4}-[(k+2)^2\pm 4(k+2)]r^{-2}
]a_{\alpha\beta}=0}
The solutions are
\eqn\threetw{a_{\alpha\beta}=C_5 v^{2}J_{(k+2\pm 2)}(v)+ C_6 v^{2}
N_{(k+2\pm 2)}(v) }

\newsec{Vector from the two-form field}

The wave equation in this case arises from \zthree\ with one free index on
 the $S^5$ and one
free index $\rho$ in the remaining 5 directions.
For $\rho=0$ we get
\eqn\sixtwo{{1\over r^3}\partial_r[r^3(\partial_r a_0+i\omega a_r)]-{(k+1)
(k+3)\over r^2}a_0=0}
For $\rho=r$ we get
\eqn\sixthree{\partial_r a_0={1\over i\omega}[\omega^2-{(k+1)(k+3)
\over r^2H}]a_r }
For $\rho=1$ we get
\eqn\sixfour{{1\over r^3}\partial_r(r^3\partial_r a_1)+[\omega^2 
H-{(k+1)(k+3)\over r^2}]a_1=0}
[The equations for $\rho=2,3$ are similar to the equation for 
$\rho=1$ so we do not need to write them.]

We find that $a_0$ and $a_r$ can be algebraically determined from 
each other. For $a_r$ we have the equation
\eqn\sixel{{1\over r}\partial_r({r\over H} a_r)+i\omega a_0=0}
Then we find the following equation for $a_r$:
\eqn\sixtw{\partial_r[{1\over r}\partial_r({r\over H} a_r)]+
[\omega^2 -{(k+1)(k+3)\over r^2H}]a_r=0}
and for $a_0$ we have the equation
\eqn\sixelqw{\partial_r[r^3\partial_r a_0(\omega^2r^2H-(k+1)(k+3))^{-1}]+a_0=0}

\subsec{Outer region}

The equations are
\eqn\sixeight{a_{1,rr}+3{a_{1,r}\over r}      +
[\omega^2 -{(k+1)(k+3)\over r^2}]a_1=0}
\eqn\sixsixt{a_{r, rr}+r^{-1} a_{r,r}+[\omega^2 -{(k+1)(k+3)\over r^2}]
a_r=0}

The solutions are
\eqn\sixnine{a_1=\tilde C_1r^{-1} J_{k+2}(\omega r) + \tilde C_2 r^{-1}
 N_{k+2}(\omega r) }
\eqn\sixsevent{a_r=C_1 J_{(k+2)}(\omega r) + C_2 N_{(k+2)}(\omega r)  }

\subsec{Intermediate region}

The equations are
\eqn\sixfive{{1\over r^3}\partial_r(r^3\partial_r a_1)-{(k+1)(k+3)\over r^2}
a_1=0}
\eqn\sixfourt{\partial_r[{1\over r}\partial_r({r({a_r\over H})})]
 -{(k+1)(k+3)\over r^2}({a_r\over H})=0}

The solutions are
\eqn\sixseven{a_1=\tilde C_3 r^{k+1} + \tilde C_4 r^{-k-3} }
\eqn\sixsevt{a_0=[C_3r^{k+1}+ C_4r^{-k-3} ] }
We find that $a_r$ vanishes if $\omega$ is set to zero, but it is still 
helpful to write it down since one can match either $a_r$ or $a_0$:
\eqn\sixfift{a_r=-(i\omega)[C_3(k+3)^{-1} Hr^{k+2} - C_4(k+1)^{-1} 
Hr^{-(k+2)}] }

\subsec{Inner region}

The equations are
\eqn\sixfourp{{1\over r^3}\partial_r(r^3\partial_r a_1)+[{\omega^2R^2\over r^4}
-{(k+1)(k+3)\over r^2}]a_1=0}
\eqn\sixtw{a_{r,rr}+9r^{-1}a_{r,r}+[{R^4\omega^2\over r^4}-
{(k+1)(k+3)\over  r^2}]a_r=0  }

The solutions are
\eqn\sixten{a_1=\tilde C_5 vJ_{(k+2)}(v) +\tilde  C_6 v N_{(k+2)}(v) }
\eqn\sixtwone{a_r = C_5 v^4 J_{(k+2)}(v) + C_6 v^4 N_{(k+2)}(v) } 
\eqn\sixtwtwo{a_0=-i\omega^2R^2[C_5(v^2J'_{(k+2)}(v)-vJ_{(k+2)}(v))
+C_6(v^2N'_{(k+2)}(v)-vN_{(k+2)}(v))]  }

\bigskip
\bigskip
\newsec{Antisymmetric tensor from the 4-form}

In \ztwo\ we take two of the free indices to be of type $\alpha\beta$ and
 the rest to be of type $\mu$. We note further that the spherical harmonics $Y^{I_{10}}_{[\alpha\beta]}$ are independent of $\partial_{[\alpha}
Y^{I_5}_{\beta]}$. Then we find the equation

\eqn\twoone{3\partial_{[\mu}a_{\nu\rho]\alpha\beta}={1\over 4}\epsilon_{\mu\nu\rho\alpha\beta}{}^{\sigma\tau\gamma\delta\epsilon}
\partial_{\gamma}a_{\delta\epsilon\sigma\tau}  }

For  $\mu,\nu,\rho=0,1,2$ we get
\eqn\twoseven{b_{3r}=\mp{\omega r H\over k+2} b_{12} }
For $\mu,\nu,\rho=r,0,3$ we get
\eqn\twofour{b_{03,r}-i\omega b_{3r}=\pm i r^{-1}(k+2)b_{12} }
For $\mu,\nu,\rho=r,1,2$ we get
\eqn\twosix{b_{12,r}=\mp ir^{-1}(k+2)b_{03} }

Eliminating $b_{03}$ and $b_{3r}$ we get
\eqn\twonine{r^{-1}\partial_r[rb_{12,r}]+[\omega^2H-{(k+2)^2\over r^2}]b_{12}=0}

\subsec{Outer region}

The equation is
\eqn\twoten{b_{12,rr}+r^{-1}b_{12,r}+[\omega^2-{(k+2)^2\over r^2}]b_{12}=0 }

 The solutions are
\eqn\twothir{b_{12}=C_1 J_{k+2}(\omega r) + C_2 N_{k+2}(\omega r) }

\subsec{Intermediate region}

 The equation is
\eqn\twoten{b_{12,rr}+r^{-1}b_{12,r}-{(k+2)^2\over r^2}b_{12}=0 }

The solutions are
\eqn\twotw{b_{12} = C_3 r^{k+2}+C_4 r^{-(k+2)} }

Thus the field equation does not have a nontrivial intermediate region.

\subsec{Inner region}

The equations are
\eqn\twoten{b_{12,rr}+r^{-1}b_{12,r}+[{\omega^2R^4\over r^{4}}-
{(k+2)^2\over r^2}]b_{12}=0 }

The solution is
\eqn\twoell{b_{12}=C_5 J_{k+2}(v) + C_6  N_{k+2}(v) }

\bigskip
\bigskip

\newsec{Two form from the antisymmetric tensor}

The wave equation follows from \zthree\ 
\eqn\fourone{\eqalign{&3{1\over \sqrt{-g}}\partial_{ \mu}[\sqrt{-g}\partial_{[\mu'}a_{\mu'_1\mu'_2]}g^{\mu'\mu}
g^{\mu'_1\mu_1}g^{\mu'_2\mu_2}]
+{1\over \sqrt{-g}}\partial_{ \alpha}[\sqrt{-g}\partial_{\alpha'}
a_{\mu'_1\mu'_2}
g^{\alpha'\alpha}g^{\mu'_1\mu_1}g^{\mu'_2\mu_2}]\cr
~~~&+2iF^{\mu_1\mu_2\mu_3\mu_4\mu_5}\partial_{\mu_3}a_{\mu_4\mu_5}=0\cr }}

Setting the free indices $\mu_1\mu_2$ to $r,3$ we get
\eqn\fourfour{a_{r3}[\omega^2H^{{1\over 2}}-{k(k+4)\over H^{{1\over
2}}r^2}]=\omega[ia_{03,r}H^{{1\over 2}}-{4R^4\over H^{{1\over 2}}r^5}a_{12}]=0}

Setting $\mu_1\mu_2$ to $1,2$ we get
\eqn\fourfive{\omega^2H^2a_{12}+{1\over r^5}\partial_r
[r^5Ha_{12,r}]-4ia_{03,r}{R^4\over r^5}
+{4R^4\omega \over r^5}a_{r3}-{k(k+4)H\over r^2}a_{12}=0}

Setting $\mu_1\mu_2$ to $0,3$ we get
\eqn\fourfivep{-{1\over r^5}\partial_r[a_{03,r}Hr^5]-{i\omega\over r^5}\partial_r[Hr^5a_{r3}]+{k(k+4)\over r^2}Ha_{03}-
{4iR^4\over r^5}a_{12,r}=0}

From the first and last equations we find
\eqn\magic{{1\over r^3}\partial_r[r^3a_{r3}]+i\omega Ha_{03}=0}

\subsec{Outer region}

In this region \magic\ becomes
\eqn\magicp{{1\over r^3}\partial_r[r^3a_{r3}]+i\omega a_{03}=0}
From this we get
\eqn\magicpp{\partial_r[{1\over r^3}\partial_r[r^3a_{r3}]]+
i\omega a_{03,r}=0}
\fourfour\ gives
\eqn\fourfourp{a_{r3}[\omega^2-{k(k+4)\over r^2}]=\omega[ia_{03,r}]}
Substituting \magicpp\ into \fourfourp\ we get 
\eqn\fourtww{a_{r3,rr}+3r^{-1}a_{r3,r}+[\omega^2-{(k(k+4)+3)\over r^2}]
a_{r3}=0  }

The solution is
\eqn\fourtwq{a_{r3}=C_1r^{-1}J_{(k+2)}(\omega r)+C_2r^{-1}N_{(k+2)}
(\omega r)}

We can then get $a_{03}$ from \magicp\ and we find
\eqn\fourth{\eqalign{&a_{03}={i\over \omega}[a_{r3,r}+3r^{-1}a_{r3}]\cr
~&
={i\over \omega}C_1[r^{-1}\partial_r J_{(k+2)}(\omega r)+2r^{-2}
J_{(k+2)}(\omega r)]+ {i\over \omega}C_2[r^{-1} \partial_rN_{(k+2)}(\omega r)+2r^{-2}N_{(k+2)}(\omega r)] \cr}}

The equation for $a_{12}$ decouples from the other fields in the
 outer region:
\eqn\fourfivep{{1\over r^5}\partial_r[r^5a_{12,r}]+\omega^2a_{12}
-{k(k+4)\over r^2}a_{12}=0}
The solution is
\eqn\fourqq{a_{12}=\tilde C_1 r^{-2}J_{(k+2)}(\omega r)+
\tilde C_2 r^{-2}N_{(k+2)}(\omega r)}

\subsec{Intermediate region}

 We get from \fourfive\ and \fourfivep\
\eqn\foursix{{1\over r^5}\partial_r[r^5Ha_{12,r}]
-{k(k+4)H\over r^2}a_{12}-4ia_{03,r}{R^4\over r^5}=0}
\eqn\fourfivep{{1\over r^5}\partial_r[a_{03,r}Hr^5]-{k(k+4)\over r^2}
Ha_{03}+{4iR^4\over r^5}a_{12,r}=0}

Define
\eqn\fourseven{\phi_\pm\equiv a_{12}\mp i a_{03} }

Then we get
\eqn\foursixp{{1\over r^5}\partial_r[r^5H\phi_{\pm,r}]
-{k(k+4)H\over r^2}\phi_{\pm} \pm {4R^4\over r^5}\phi_{\pm,r}=0}

The solutions are
\eqn\foureight{\phi_-= C_3^-r^{-k}[r^4+R^4]^{-1}+C_4^-r^{k+4}
[r^4+R^4]^{-1} }
\eqn\fournine{\phi_+=C_3^+r^k+C_4^+r^{-(k+4)} }

\subsec{Inner region}

Here we use the factorized form of the equations. In this region 
the equation can be written as
\eqn\fourqone{[{2k\over R}+i*D][{2(k+4)\over R}-i*D]a_{\mu\nu}=0}
\eqn\fourqtwo{(*Da)_{\mu\nu}=-\epsilon_{\mu\nu}{}^{\mu_1\mu_2\mu_3}
\partial_{\mu_1}a_{\mu_2\mu_3} }
where
\eqn\zm{\epsilon_{12r03}=
{r^3\over R^3}  }
sets the sign  of the volume form in the $AdS_5$.

Assume that the second factor vanishes; the analysis is similar when the
 first factor vanishes. Then
we get \eqn\fourqfive{[{2(k+4)\over R}-i*D]a_{\mu\nu}=0}
which reads
\eqn\fourqfivep{{2(k+4)\over R}a_{\mu\nu}+i\epsilon_{\mu\nu}{}^
{\mu_1\mu_2\mu_3}\partial_{\mu_1}
a_{\mu_2\mu_3} =0  }
The $r3$ component is
\eqn\fourqsix{{(k+4)\over R}a_{r3}+{\omega R^3\over r^3}a_{12}=0}
The $12$ component is
\eqn\fourqseven{{(k+4)\over R}a_{12}+{\omega r\over R}a_{r3}-
{ir\over R}a_{03,r}=0}
The $03$ component is
\eqn\fourqeight{{(k+4)\over R}a_{03}+i{r\over R}a_{12,r}=0}

From the $03$ equation we find $a_{03}$ and substitute it into the 
$12$ equation. Then we get
\eqn\fourqten{{1\over r}\partial_r[ra_{12,r}]-{(k+4)^2\over r^2}
a_{12}+{\omega^2R^4\over r^4}a_{12}=0 }

The solution is
\eqn\fourqel{a_{12}=C_5 J_{(k+4)}(v)+C_6N_{(k+4)}(v) }

From this we then get
\eqn\fourqtw{a_{r3}=-{\omega R^4\over r^3(k+4)}[C_5 J_{(k+4)}(v)+
C_6N_{(k+4)}(v)]}
\eqn\fourqthir{a_{03}=-{ir\over (k+4)}[C_5 \partial_rJ_{(k+4)}(v)+
C_6\partial_rN_{(k+4)}(v)]}

If in \fourqone\ we assume that the first factor vanishes 
(The two factors commute in the
inner region) we get the solutions
\eqn\fourqel{a_{12}=C_5 J_{(k)}(v)+C_6N_{(k)}(v) }
\eqn\fourqtw{a_{r3}={\omega R^4\over r^3 k}[C_5 J_{(k)}(v)+C_6N_{(k)}(v)]}
\eqn\fourqthir{a_{03}={ir\over k}[C_5 \partial_rJ_{(k)}(v)+
C_6\partial_rN_{(k)}(v)]}

\subsec{The case $k=0$.}

The case with $k=0$ is special because in this case we have a 
gauge freedom given by a function $\Lambda_\mu(x)$ which was not
 fixed by the choices \zc , \zb\ \van . This case was studied in
 \ref\raj{Arvind Rajaraman, `Two-Form Fields and the Gauge Theory
 Description of
Black Holes', hep--th/9803082}. To make the solutions 
\foureight , \fournine\ in the intermediate region agree
 with the corresponding solutions given in this reference, we have to 
note that this gauge freedom allows us to add arbitrary constants to 
$a_{12}$ and $a_{03}$.
 This freedom removes two of the four constants that appeared 
in the solution for general $k$, and corresponds to the fact that
 only one propagating degree of freedom 
exists for $k=0$ in contrast to two degrees of freedom for higher $k$. 
Thus we can solve the dynamical equations for $a_{12}$ and then note that
 for $k=0$ we can find $a_{03}$ algebraically from the equations of motion.

\newsec{Absorption probabilities, structure of solutions}

First we consider the fields for which the solution in the intermediate 
region is a pure power law.
In these cases we can ignore the intermediate region in the calculation 
of scattering at leading
order in $\omega$, since the solution form in the outer part of the inner
 region directly joins up
with the solution in the inner part of the outer region. The fields in 
this category are

(a)\quad The dilaton-axion scalar $B$

(b)\quad The two form field in the $x$ space arising from the 4-form 
field $A_{\mu\nu\alpha\beta}$ 

(c) \quad The vector $A_\mu$ in the $x$ space arising from the two form field
$A_{\mu\alpha}$. Here we have to study the transverse components $a_a, a=1,2,3$
and the longitudinal component described by $a_r, a_0$; both have trivial 
intermediate region
solutions.

For all these fields the absorption probability \zg\ is 
\eqn\xq{{\cal P}=4\pi^2({\omega R\over 2})^{4k+8}
[\Gamma(k+2)\Gamma(k+3)]^{-2} }

Now we consider the scalar in the $x$ space arising  from the 
2-form field $A_{\alpha\beta}$. The
spherical harmonics for this case split into two classes denoted
 by $Y^{k,\pm}_{[\alpha\beta]}$. The
absorption probabilities for these two classes are given by
\eqn\fiveel{{\cal P}_+=4\pi^2({\omega R\over 2})^{4k+12}
[\Gamma(k+3)\Gamma(k+4)]^{-2} }
\eqn\fiveelp{{\cal P}_-=4\pi^2({\omega R\over 2})^{4k+4}
[\Gamma(k+1)\Gamma(k+2)]^{-2} }

For the 2-form field in the $x$ space arising from $A_{\mu\nu}$ we
 have seen that
the longitudinal and transverse components do not decouple, except 
for $k=0$ when the longitudinal
component is not dynamical. It appears to not be helpful to try to
 define cross sections in this
case, so we present instead the relation between the solutions at
 $r\rightarrow 0$ and the
solutions at $r\rightarrow\infty$.

In the outer region we have found in \fourqq\fourtwq\fourth\
\eqn\fourqqp{a_{12}=\tilde C_1 r^{-2}J_{(k+2)}(\omega r)+
\tilde C_2 r^{-2}N_{(k+2)}(\omega r)}
\eqn\fourtwqp{a_{r3}=C_1r^{-1}J_{(k+2)}(\omega r)+C_2r^{-1}
N_{(k+2)}(\omega r)}
\eqn\fourthp{a_{03}={i\over \omega}[a_{r3,r}+3r^{-1}a_{r3}] }

In Table 1 we present the result of matching the solutions across 
the intermediate region.
The first column gives the choice of solution at $r\rightarrow 0$.
 The other colums give the
solution at $r\rightarrow\infty$ in terms of the  coefficients in
 \fourtwqp , \fourthp .

\midinsert
{\localmag{850}{85}{100} 
$$
\vbox{\offinterlineskip
\def\strut{\vrule height 3.25ex  width 0pt depth 2ex}
\def\hline{\noalign{\hrule}}
\halign{\vrule#\hfil\strut&\quad#\hfil\quad\vrule&&\quad\enspace
\hfil#\hfil\quad\vrule\cr
\noalign{\hrule}
&$\hfil a_{12}$&$\tilde C_1$&$\tilde C_2$&$C_1$&$C_2$\cr
\noalign{\hrule}
&$J_{(k+4)}(v)$&$0$&
$-{\pi R^2  \over \Gamma(k+2)\Gamma(k+5) }({\omega R\over 2})^{2k+6}$
&$0$&$-{ \omega\over k}\tilde C_2$\cr
\hline
&$N_{(k)}(v)$&$-{  R^2 \Gamma(k)\Gamma(k+3)\over \pi }({\omega R\over 2})^{-(2k+2)}$&$0$&${\omega\over k+4}   \tilde C_1$&$0$\cr
\hline
&$N_{(k+4)}(v)$&$-{ R^2 \Gamma(k+4)\Gamma(k+3)\over  \pi }({\omega R\over 2})^{-(2k+6)}$&$0$&$-{\omega\over k+4}   \tilde C_1$&$0$\cr
\hline
&$J_{(k)}(v)$&$0$&
$-{\pi  R^2\over \Gamma(k+1)\Gamma(k+2) }({\omega R\over 2})^
{2k+2}$&$0$&${ \omega\over k}\tilde C_2 $\cr
\hline
}}
$$ }
{\narrower Table 1: Scattering matrix for the two form from the
 antisymmetric tensor\par}
\endinsert

In the first two cases the solution in the intermediate region is 
a pure power law, as it
gives $\phi_+$ (eq. \fournine\ ) in this region. In the last two 
cases the solution is not trivial in
the intermediate region, since it gives $\phi_-$ (eq. \foureight\ )

\newsec{Analysis of solutions}

Let us  comment on the nature of the calculation and its results. 
We note that in the outer and  the inner regions  the waveform 
attains the form of a freely traveling wave. The intermediate 
region acts as a barrier that the solution has to tunnel through,
 and at small $\omega$ the tunneling is small. We can imagine 
having the wave incident from either side of this barrier. If we 
throw in a particle from infinity and have it absorbed by the brane,
 then the particle is incident from the outer region, and the boundary
 condition on the solution is that there be no outgoing wave at 
$r\rightarrow 0$ in the inner region. Perhaps for emission of a 
quantum from the branes we should take the particle to be incident 
from the inner region.

The results on the absorption probabilities are summarized in Table 2.
 The first column gives the the field and its expansion in spherical 
harmonics. The second column gives the range of the index  for 
the harmonics. The third column lists the cases that occur in the
 supergravity multiplet of the 5-dimensional AdS space. 
(These fields are labelled by $\odot$ in \van , and we have
 used the same notation here.) The fourth column gives the 
absorption probabilities. For the antisymmetric tensor with 
no indices along the $S^5$ the components mix in a way that 
does not make it helpful to define an absorption probability.
 One can choose combinations of fields incident on the branes
 such that the reflected fields are a multiple of the combination
 that was incident. In this case one can define an absorption 
probability for the incident combination, but it turns out that
 in this basis the two power laws of $\omega$ that arise during 
the calculation mix with each other, so the diagonalization 
obscures the $\omega$ dependence of the solution.

For the case $k=0$ however we have seen that only one degree of
 freedom propagates, and the absorption probability can be computed
for this mode. The result from \raj\ can be written as
\eqn\rajj{{\cal P}_-={4\pi^2\over [\Gamma(k+3)\Gamma(k+4)]^2} 
({\omega R\over 2})^{4k+12}}
which agrees with the general pattern of results in Table 1.

In the case of the scalar field $B$ the above computation was 
carried out for all partial waves in \kleb\gubs , and our result
 for this agrees with the results presented there. 

The probability for passage from one side of the barrier to the
 other is the same, whichever side the particle is incident from,
 as would be expected on general grounds. 
In the intermediate region (the barrier) 
we set $\omega=0$. There are two solutions to a second order wave 
equation, and these solutions are power laws in $r$ in the inner 
part of the intermediate region, and also in the outer part of the
 intermediate region. The powers are not the same in these two parts
 of the intermediate region in general, since the solutions need 
not be a simple power of $r$ throughout the region. Let the powers
 in the inner part of the intermediate region be
\eqn\eighttwo{r^{a+b}, ~~~r^{-a+b}   }
Let the powers in the outer part of the intermediate region be
\eqn\eightthree{r^{c+d}, ~~r^{-c+d} }
Here we have  assumed that $a\ge 0, ~c\ge 0$.
Then the absorption probability has the dependence
\eqn\eightone{{\cal P}\sim \omega^{2a+2c} }
The parameters $a$ and $c$ are the indices of the Bessel function $J$
in the inner and outer regions respectively.

We summarize some information about the solutions in Table 3. 
In the first column we list the fields that we study. Here the symbols 
$a_+, a_-$ correspond to the two kinds of spherical harmonics that we
 have for this field,  $Y^{k,+}_{[\alpha\beta]}$ and 
$Y^{k,-}_{[\alpha\beta]}$ respectively. The second column gives the two
 powers of $r$ for each field in the inner part of the intermediate region, 
eq. \eighttwo . The third column gives the powers of $r$ in the outer part
 of the intermediate region, eq. \eightthree .

The fourth column gives the dimension $\Delta$ of the field in the Yang-Mills
 theory which corresponds to the mode of supergravity having the given 
power law behaviors in the inner region. This dimension is given by 
\witten\ref\oz{Yaron Oz and John Terning, `Orbifolds of $AdS_5\times
 S^5$ and 4d Conformal
Field Theories', hep-th/9803167}
\eqn\zn{\tilde m^2=(\Delta-p)(\Delta+p-4)}
where $\tilde m^2$ is the eigenvalue of the Maxwell operator on the
 appropriate p-form in the 5-dimensional AdS geometry in the inner 
region. These $\tilde m^2$ values are also the same as those listed 
in \van . Note that $\Delta$ is connected to the power $a+b$ of  $r$ 
in the solution of the field equation in the outer part of the inner
 region \witten ; this is the power that decays as we go towards 
smaller $r$ values. The relation is
\eqn\zv{\Delta = a+b-p+4}
Where $p$ is the rank of the form in the AdS space.

The last column gives the $\omega$ dependence of the absorption 
probability, which is given by \eightone . For the antisymmetric
 tensor with no indices along the $S^5$ the components mix  
and we do not get a statement analogous to \eightone . For 
the case $k=0$,  
the absorption probability \rajj\ agrees with the pattern of
 results in Table 2 for the $\Delta=6$.

\midinsert
$$
\vbox{\offinterlineskip
\def\strut{\vrule height 3.25ex  width 0pt depth 2ex}
\def\hline{\noalign{\hrule}}
\halign{\vrule#\hfil\strut&\quad#\hfil\quad\vrule&&\quad\enspace
\hfil#\hfil\quad\vrule\cr
\noalign{\hrule}
&\hfil {Field}&$k$&$\odot$&${\cal P}$\cr
\noalign{\hrule}
&$B= B^{I_1}Y^{I_1}$&$k\geq 0$&$k=0$&${\cal P}= 
{4\pi^2\over [\Gamma(k+2)\Gamma(k+3)]^2} ({\omega R\over 2})^{4k+8}$\cr
\hline
&$A_{\alpha\beta}= a^{I_{10,+}}Y^{I_{10,+}}_{[\alpha\beta]}$
&$k\geq 1$&&$
{\cal P}_-={4\pi^2\over [\Gamma(k+3)\Gamma(k+4)]^2} ({\omega R\over 2})
^{4k+12}$
\cr 
&$A_{\alpha\beta}= a^{I_{10,-}}Y^{I_{10,-}}_{[\alpha\beta]}$&$k\geq 
1$&$k=1$&${\cal P}_+=
{4\pi^2\over [\Gamma(k+1)\Gamma(k+2)]^2} ({\omega R\over 2})^{4k+4}$
\cr
\hline
&$A_{\mu\alpha}=a_\mu^{I_{5}}
Y^{I_{5}}_\alpha$&$k\geq 1$&&${\cal P}=
{4\pi^2\over [\Gamma(k+2)\Gamma(k+3)]^2} ({\omega R\over 2})^{4k+8}$\cr
\hline
&$a_{\mu\nu\alpha\beta}=
b_{\mu\nu}^{I_{10,\pm}}Y^{I_{10,\pm}}_{[\alpha\beta]}$&$k\geq 1$
&&${\cal P}=
{4\pi^2\over [\Gamma(k+2)\Gamma(k+3)]^2} ({\omega R\over 2})^{4k+8}$\cr
\hline
&$A_{\mu\nu}= a_{\mu\nu}^{I_{1}}Y^{I_{1}}$&$k\geq 1$&$k=1$&
${\scriptstyle\rm   }$\cr
&&$k\geq 0$&&${\scriptstyle\rm    }$\cr
\noalign{\hrule}
}}
$$
{\narrower Table 2: Fields and corresponding absorption probabilities.
\par}
\endinsert

\midinsert
$$
\vbox{\offinterlineskip
\def\strut{\vrule height 3.25ex  width 0pt depth 2ex}
\def\hline{\noalign{\hrule}}
\halign{\vrule#\hfil\strut&\quad#\hfil\quad\vrule&&\quad\enspace
\hfil#\hfil\enspace
&\enspace\hfil#\hfil\quad\vrule\cr
\noalign{\hrule}
&\hfil {Field}&\multispan2 \hfil Inner powers\hfil \vrule &\multispan2
 \hfil Outer powers\hfil\vrule&\omit
\hfil\qquad $\Delta$\hfil\qquad \vrule &${\cal P}$\cr
\noalign{\hrule}
&$~~B$& \hfil $r^k$ & $r^{-k-4}$&$r^k,$&$r^{-k-4}$&\omit\hfil $k+4$\hfil
\vrule&$\omega^{4k+8}$\cr
\hline
&$~~a^+$&$r^{k+2},$&$r^{-k-6}$&$r^{k+2},$&$r^{-k-2}$&\omit\hfil$k+6$\hfil \vrule&$\omega^{4k+12}$\cr 
&$~~a^-$&$r^{k-2},$&$r^{-k-2}$&$r^{k+2},$&$r^{-k-2}$&\omit\hfil$k+2$\hfil
 \vrule&$\omega^{4k+4}$\cr
\hline
&$~~a_1$&$r^{k+1},$&$r^{-k-3}$&$r^{k+1},$&$r^{-k-3}$&\omit\hfil$k+4$\hfil 
\vrule&$\omega^{4k+8}$\cr
&$~~a_0$&$r^{k+1},$&$r^{-k-3}$&$r^{k+2},$&$r^{-k-2}$&\omit\hfil$k+4$\hfil
 \vrule&$\omega^{4k+8}$\cr
\hline
&$~~b_{\mu\nu}$&$r^{k+2},$&$r^{-k-2}$&$r^{k+2},$&$r^{-k-2}$&\omit\hfil$k+4$
\hfil \vrule&$\omega^{4k+8}$\cr
\hline
&$a_{12}-ia_{03}$&$r^k,$&$r^{-k-4}$&$r^k,$&$r^{-k-4}$&\omit\hfil$k+2$\hfil 
\vrule&
${\scriptstyle\rm    }$\cr
&$a_{12}+ia_{03}$&$r^{k+4},$&$r^{-k}$&$r^k,$&$r^{-k-4}$&\omit\hfil$k+6$\hfil
 \vrule&
${\scriptstyle\rm    }$\cr
\noalign{\hrule}
}}
$$
{\narrower Table 3: Power law behavior for the fields in the inner and outer
 regions, and the dependence of absorption probability on $\omega$. \par}
\endinsert

\newsec{Discussion}

In the geometry produced by the 3-branes we have the asymptotic 
regions of large $r$ and small $r$, in each of which we have the
 maximal number of supersymmetries (32 real components). In the 
intermediate region the number of supersymmetries is only half 
that (16 real). Consider any partial wave,  and  imagine that 
it is incident from the large $r$ region.  In passing through 
the intermediate region there will be a change in the fields, 
such that a possibly different basis of components will emerge
 as the natural one in the $r\rightarrow 0$ region.

It would be interesting to seek a microscopic model of the 3-branes
 which reflects the action of the intermediate region. What appears
 to be passage into the inner region in the classical picture is 
expected to be an excitation of the quantum states on the branes 
in a microscopic description. Note that we have considered linear
 perturbations on the background, but in general the strength of 
the perturbation grows as the wave moves deeper into the inner region.
(This phenomenon occurs also for perturbations on the three 
charge hole in 4+1 dimensions.)

Recently it has been shown that in the three charge black hole 
(in the dilute gas regime) the  two point function of supergravity
 fields in the AdS region  can be used to reproduce the agreement
 between the microscopic and classical absorption cross sections, 
for s-wave absorption \ref\teo{E. Teo, `Black hole absorption
 cross-sections and the anti-de Sitter- conformal field theory 
correspondence', hep--th/9805014.}. It is important in this calculation 
to use the correctly normalized two-point function for the field 
\polyakov\ref\freed{D.Z. Freedman, S.D. Mathur, A. Matusis, and L.
 Rastelli, `Correlation functions in the $CFT(d)/AdS_{d+1}$ correspondence',
 hep--th/9804058.}. It is not clear however how such a calculation would be 
extended to higher partial waves.  From Table 2 we observe that if
 $\Delta=k+m$, then the power of $\omega$ is $4k+2m$. Thus the case 
$k=0$ might not exhibit the full structure of the relation between 
the classical calculation and the microscopic theory. We hope to 
return to this issue in the future.

\bigskip
\bigskip
\bigskip

\centerline{{\bf Acknowledgements}}

We would like to thank for discussions D. Freedman, I. Klebanov, 
P. van Nieuwenhuizen, L. Rastelli and  S.P.Trivedi. 
We are especially grateful to S.R. Das for many insights. This work
 is partially supported by cooperative
agreement number DE-FC02-94ER40818.
 
\vfill
\eject

\listrefs

\bye

\end